\documentclass[printer]{aa}
\usepackage[english]{babel}
\usepackage{graphicx}
\usepackage{newlfont}
\usepackage{amssymb}
\usepackage{subfig}
\usepackage{commath}

\usepackage{amsmath}
\usepackage{latexsym}
\usepackage{amsthm}
\usepackage{natbib}
\usepackage{microtype}
\bibpunct{(}{)}{;}{a}{}{,} 
\usepackage{longtable}
\usepackage{rotating}

\usepackage[usenames,dvipsnames]{color}
\usepackage[bookmarks, colorlinks, breaklinks]{hyperref}  
\hypersetup{linkcolor=blue,citecolor=MidnightBlue,filecolor=black,urlcolor=MidnightBlue} 

\usepackage[varg]{txfonts}

\newcommand{\figiac}[1]{Fig. \ref{#1}}

\newfont{\gwpfont}{cmssq8 scaled 1000}

\def\YX {Y_{\textrm X}}
\def\YSZ {Y_{\textrm SZ}}

\def\Mv {M_{500}}
\def\Mv {M_{500}}
\def\Rv {R_{500}}
\def\Rvyx {R_{500}^\mathrm{Y_{X}}}
\def\Mvyx {M_{500}^\mathrm{Y_{X}}}

\def\Mvher500{M^\mathrm{HE}}

\def\MY {$M_{500}$--$Y_{\textrm X}$}
\def\MYSZ {$\Mv$--$\YSZ$}

\def\ammasso{ClG-J104803.7+313843}

\def\xmm{XMM-{\it Newton}}
\def\planck{{\it Planck}}

\def\cor

\def\msol{$[10^{14} M_{\odot}]$}

\begin{document}

\title{X-ray characterisation of the massive galaxy cluster ClG-J104803.7+313843 at z=0.76 with XMM-\textit{Newton}  }

\author{I. Bartalucci\inst{1} \and F. Gastaldello\inst{1} \and E. Piconcelli\inst{3} \and L. Zappacosta\inst{3} \and M. Rossetti\inst{1} \and S. Ghizzardi\inst{1} \and S. De Grandi\inst{2} \and S. Molendi\inst{1} \and Marco Laurenti\inst{3,4}} 

 \institute{INAF - Istituto di Astrofisica Spaziale e Fisica Cosmica di Milano, Via A. Corti 12, 20133 Milano, Italy
         \and
         INAF - Osservatorio Astronomico di Brera, via E. Bianchi 46, 23807 Merate (LC), Italy
         \and
         INAF - Osservatorio Astronomico di Roma, via di Frascati 33, 00078 Monte Porzio Catone, Italy
         \and
         Dipartimento di Fisica, Università di Roma “Tor Vergata”, Via della Ricerca Scientifica 1, I-00133 Roma, Italy
}
\date{}

\abstract{We present the characterisation of the massive cluster \ammasso\ at z=0.76 performed using a serendipitous \xmm\ observation. 
High redshift and massive objects represent an ideal laboratory to benchmark our understanding of how cluster form and assembly formation driven mainly by gravity. 

Leveraging the high throughput of \xmm\ we were firstly able to determine the redshift of the object, shedding light on ambiguous photometric redshift associations. 
We investigated the morphology of this cluster which shows signs of merging activities in the outskirts and a flat core. We also measured the radial density profile up to $\Rv$. With these quantities in hand, 
we were able to determine the mass, $\Mv = 5.64^{+0.79}_{-0.62} \times 10^{14} M_{\odot}$, using the $\YX$ proxy. This quantity improves previous measurement of the mass of this object by a factor of $\sim 3.5$.
The characterisation of one cluster at such mass and redshift regime is fundamental as these objects are intrinsically rare, the number of objects discovered so far being less than $\sim 25$. Our study highlights the importance of using X-ray observations in combination with ancillary multi-wavelength data to improve our understanding of  high-z and massive  clusters}
 \titlerunning{---}
\authorrunning{Bartalucci et al.}

\maketitle

\section{Introduction}\label{sec:introduction}
Galaxy clusters are fundamental tools to test the standard $\Lambda$Cold-Dark-Matter($\Lambda$CDM) paradigm for structure formation. 
Their abundance as a function of time is sensitive to the underlying cosmology (e.g. \citealt{vikh2009,allen2011}). Furthermore, the dark matter
(DM) density profile shape probes the gravitational collapse history. In the standard scenario the collapse of
DM is gravity driven and hence scale-free. 

The success of galaxy clusters in helping establishing the current understanding of the Universe, from the existence and nature of DM (\citealt{zwicky33} and \citealt{clowe06}) going through the ruling out of cosmological models, as the model with a critical matter density \cite{white93}, and to the constraints to the $\Lambda$CDM model (e.g. \citealt{allen2004}, \citealt{vikh2009}, and \citealt{mantz10}) have been based upon observations of the most massive clusters, $M_{500} > 5 \times 10^{14} M_{\odot}$\footnote{ $R_{\Delta}$ is defined as the radius enclosing $\Delta$ times the critical matter density at the cluster redshift. $M_{\Delta}$ is the corresponding mass.}, at relatively low redshifts where well characterised samples with high quality observations exist.
The investigation of massive clusters at high redshifts has key potential to progress our understanding. Within the cosmological context, the sensitivity of the evolution of the cluster mass function is enhanced at the high mass end. Furthermore, the abundance of the most extreme massive clusters is sensitive to the details of the initial fluctuations from inflation, such as the existence of high mass, high redshift clusters can be used to identify deviations from $\Lambda$CDM (e.g. \citealt{harrison11} and \citealt{harrison13})

\begin{figure*}[!ht]
\begin{center}
\includegraphics[width=0.9\textwidth]{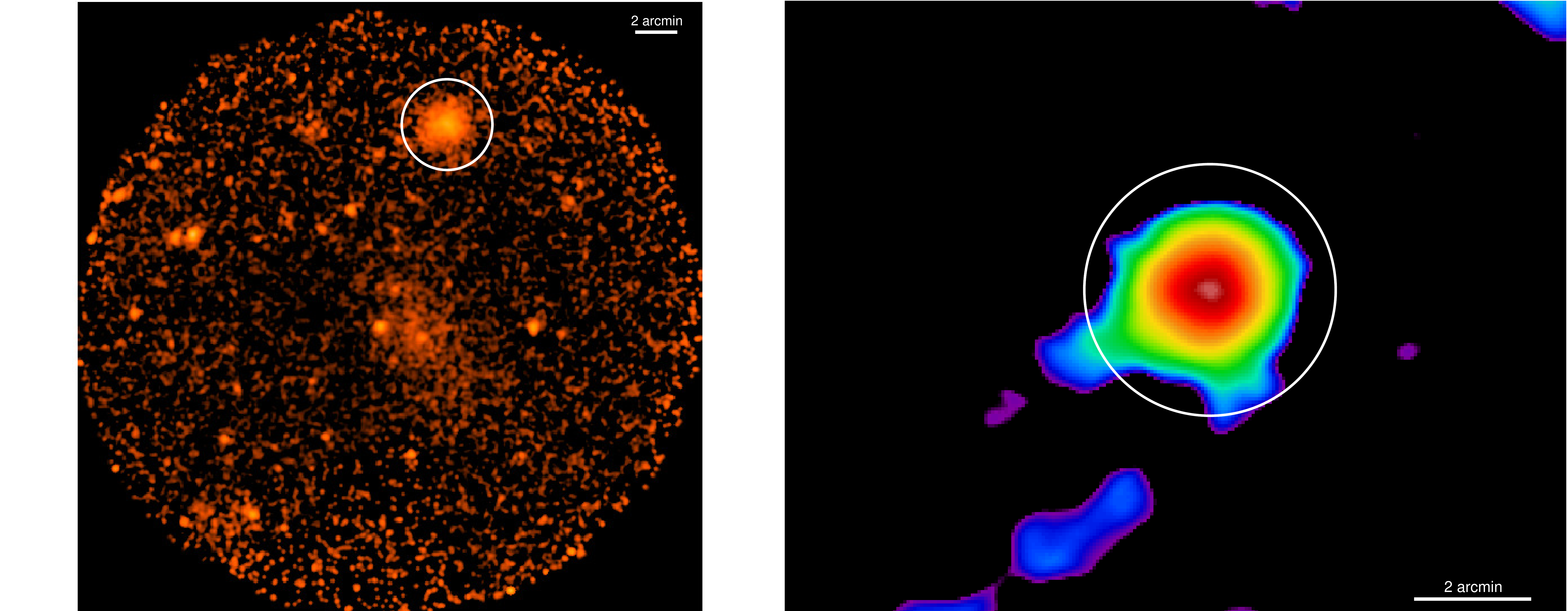}
\end{center}
\caption{\footnotesize{\textit{Left panel:} Background-subtracted and exposure-corrected image in the $[0.5-2.5]$ keV band of the full field of view of the \xmm\ observation used in this work. This image is obtained combining the 3 EPIC cameras. The cluster is in the North-West sector. The white circle encompasses $R_{500}$ derived using the $Y_X$ proxy, assuming self-similar evolution, as described in Section \ref{sec:global}. \textit{Right panel:} Wavelet filtered image of the cluster in the $[0.5-2.5]$ keV band. The emission shown is detected at a significance level greater than $3\sigma$. The white circle is the same as in the left panel.  The colour bar refers to the count level of the wavelet filtered image}}
\label{fig:clus_image}
\end{figure*}

Recent works obtained surprising results using high redshift objects. \cite{macdonald2017} has shown the remarkable stability of
cool core clusters from stacking a sample of 139 clusters in the redshift range $0.025 < z < 1.2$ in 5 redshift bins. On the simulations side, \cite{lebrun18} studied the evolution of the DM profiles of the most massive clusters, $M > 5.5 \times 10^{14} M_{\odot}$ , extracted from a large suite of cosmological simulations and found little evolution with redshift. \cite{bartalucci18} started to test these predictions by measuring the hydrostatic mass of 5 massive objects at z$\sim 0.9$ out to $\Rv{}$. 
Unfortunately, high redshift and massive objects are intrinsically rare. Furthermore, X-ray observations are extremely challenging because of the cosmological dimming. 

The importance of the difficult task to detect and characterise clusters has motivated a substantial effort at various wavelengths: through X-rays with ROSAT (e.g. \citealt{rosati98} and \citealt{ebeling01}) and \xmm\ (e.g. \citealt{fassbender11} and \citealt{willis13}), through optical and infrared data (e.g. SPARCS:  \citealt{muzzin09} and MADCoWS: \citealt{gonzalez19}) and through the Sunyaev-Zel’Dovich effect (SZ, \citealt{sunyaev1980}) with large portion of the sky surveys such as the \textit{Planck} all sky-survey \citep{planck_esz,planckpsz1,planck_psz2}, the South Pole Telescope survey (SPT, \citealt{bleem2015}), or the Atacama Cosmology Telescope (ACT, \citealt{hasselfield2013,marriage2011}).
The leverage of these objects for astrophysical as well as cosmological purposes requires X-ray and optical follow-ups. Generally speaking, their fundamental quantities such as $\Mv$ or the redshift are affected by large uncertainties. Ideally, X-ray deep observations are required to obtain thermodynamic and dynamic radial profiles and fully exploit clusters as cosmological probe (e.g. see \citealt{bartalucci17,bartalucci18,bartalucci19}). Such observations are extremely time demanding and the construction of a sample of objects whose global quantities are well characterised is fundamental to carefully pick the objects. 

In this context, we present the X-ray analysis of the cluster ClG-J104803.7+313843. This object is part of a sample of 44 candidate clusters presented in \cite{clus_candiate15} which have been confirmed by an optical follow-up using the William Herschel Telescope (WHT) and the Large Binocular Telescope (LBT). The authors used the optical datasets to measure the redshift and the richness of these clusters, which were initially detected combining RASS and SDSS datasets. Furthermore, the authors analysed the SZ CARMA observations for 21 clusters, finding an SZ signature for 11 of them.
The optical photometric redshift of \ammasso\ is 0.75 and its mass estimated via the $M-Y_{SZ}$ relation from the CARMA dataset is $M_{500}=9.8\pm 3.2 \times 10^{14} M_{\odot}$. This objects falls within the field of view of the \xmm\ observation ID 0843830401 targeting the AGN J104817.98+312905.8. Interestingly, this cluster has not been found by the full sky survey of \textit{Planck} neither in any previous X-ray catalogue, but it appears as candidate in the Combined Planck-RASS catalogue of X-ray-SZ clusters published by \cite{paula19} at $z=0.5$.

The paper is organised as follows: the data preparation is presented in Section 2, we present the analysis of the cluster and the results in Section 3, and we discuss the results in Section 4.
We adopt a flat $\Lambda$-cold dark matter cosmology with $\Omega_m = 0.3$, $\Omega_\Lambda = 0.7$, $H_{0} = 70$ km Mpc s$^{-1}$, and  $h(z) = (\Omega_m (1+z)^3 + \Omega_\Lambda)^{1/2}$ throughout. Uncertainties are given at the 68 \% confidence level ($1\sigma$). All fits were performed via $\chi^2$ minimisation. 

\section{Data Preparation}\label{sec:data_sample}

The \xmm\ observation ID 0843830401 (PI: Piconcelli) was taken using the European Photon Imaging Camera (EPIC, \citealt{struder2001,turner2001}). The camera is formed by three detectors MOS1, MOS2 and \textit{pn} that observe simultaneously the same object. This observation has an exposure time of 18 ks. We follow the reduction procedure detailed in section 2.3 of \cite{bartalucci19} and report here briefly the main steps. The dataset has been reduced by applying the latest calibration files using the Science Analysis
System (SAS)\footnote{\url{cosmos.esa.int/web/xmm-newton}} pipeline version 18.0 and calibration files as available in July 2021 by using \textit{emchain} and \textit{epchain} tools. Events for which the PATTERN keyword is $>4$ and $>13$ for
MOS1,2, and \textit{pn} cameras, respectively, were removed from the analysis.
We filter the datasets from flares and we obtain an useful exposure time of 16.3 ks and 12.6 ks for MOS1,2 and PN cameras, respectively. 
Exposure maps were computed using the SAS tool \textit{eexpmap} and the vignetting is taken in account following the weighting scheme of \cite{arnaud2001} and using the SAS tool \textit{evigweight}.
At the end of these steps, we combine the datasets from the three detectors to maximise the statistic.
We identified point sources using the Multi-resolution wavelet software \cite{sta98} and masked out them from the analysis.

X-ray observations are affected by the sky and the instrumental background. The latter component is formed by the interaction of high energy particles with the detectors and is removed following the procedure described in Section 3.1 of \cite{bartalucci18}. Briefly, we subtract the particle background by using tailored instrumental background datasets.
The sky background that affects X-ray observations is removed differently for the imaging i.e. surface brightness profile and spectroscopy analysis. For this reason we explain these procedures in Section 3. This component is formed by the Galaxy thermal emission and the superimposed emission of all the
unresolved point sources, namely the cosmic X-ray background (\citealt{lumb2002,kuntzsnowden2000,giacconi2001}). 

We also performed the wavelet filtering of the exposure-corrected and background subtracted image in the $[0.5-2.5]$ keV band using a soft $3\sigma$ thresholding of B3-spline wavelet coefficients, the significance thresholds being computed from Poissonian realisations of the image, following the stabilisation scheme of \cite{zhang08} and the procedures described in \cite{bourdin08}. The wavelet filtered image is shown in the right panel of \figiac{fig:clus_image}.

\section{Cluster analysis}\label{sec:clus_prop}
\subsection{Morphology}
We show in the left panel of \figiac{fig:clus_image} the exposure corrected and background subtracted image of the field of view containing the cluster \ammasso. The object is located in the North-West sector and is $\sim 9$ arcmin off-axis from the centre of the observation. We firstly determined the position of the X-ray peak by determining the maximum in the count-rate image in the $[0.3-2.5]$ keV band
after being smoothed using a 2-dimensional Gaussian kernel with a width of 4 pixels. The coordinates are reported in Table
\ref{tab:clus_prop}. 

We the results of the wavelet filtered map in the right panel of \figiac{fig:clus_image}. There are two behaviours regarding the morphology of the cluster. 
The inner part of the object within $\sim 1.5$ arcmin appears to be quite regular showing a roundish shape with a moderately bright core. We used the ratio of the flux computed within fixed apertures, $C_{SB}$, to measure the concentration of the cluster using the technique described in Section 4.2 of \cite{bartalucci19} and we obtain $C_{SB}=0.22 \pm 0.02$. The PSF is accounted in the calculation using the model of \cite{ghizzardi2001}.
Generally speaking, clusters are considered to be concentrated and possible candidates to host a cool-core if the $C_{SB}>0.3$ (e.g. see \citealt{bartalucci19} and references therein). The morphology appears to be more irregular at large scales, the shape being ellipsoidal and elongated along the NW-SE direction. There are two faint substructures appearing in the S and SE sectors. The X-ray morphology is consistent with the SZ morphology of the CARMA data shown in Fig. D1 of \cite{clus_candiate15}.

\subsection{Redshift confirmation}\label{sec:redshift}
The redshift of a cluster can be determined in X-ray measuring the shift of the 7 keV iron line, successfully performed in e.g. \cite{yu11} and \cite{planck_xmm_plckg266}. Generally speaking, this measurement is particularly challenging because at such energy the effective area of X-ray telescope is particularly low. However, the effective energy of this line for a high-redshift object is moved towards lower energies where the effective area is significantly higher and thus the statistic can be sufficient to determine the position of the line and thus the redshift. This offers and unique opportunity to confirm the redshift of \ammasso\ and to benchmark the X-ray capabilities.

The spectral analysis to determine the redshift is as follows. We defined a circular region centred on the X-ray peak whose radius is defined to maximise the SNR. We extracted the spectrum from each of the three detectors following the procedures described in \cite{pra10} and Section 3.4 of \cite{bartalucci17}. The spectra are subtracted from the instrumental background using the tailored background datasets.
\begin{figure}[!ht]
\begin{center}
\includegraphics[width=1\columnwidth]{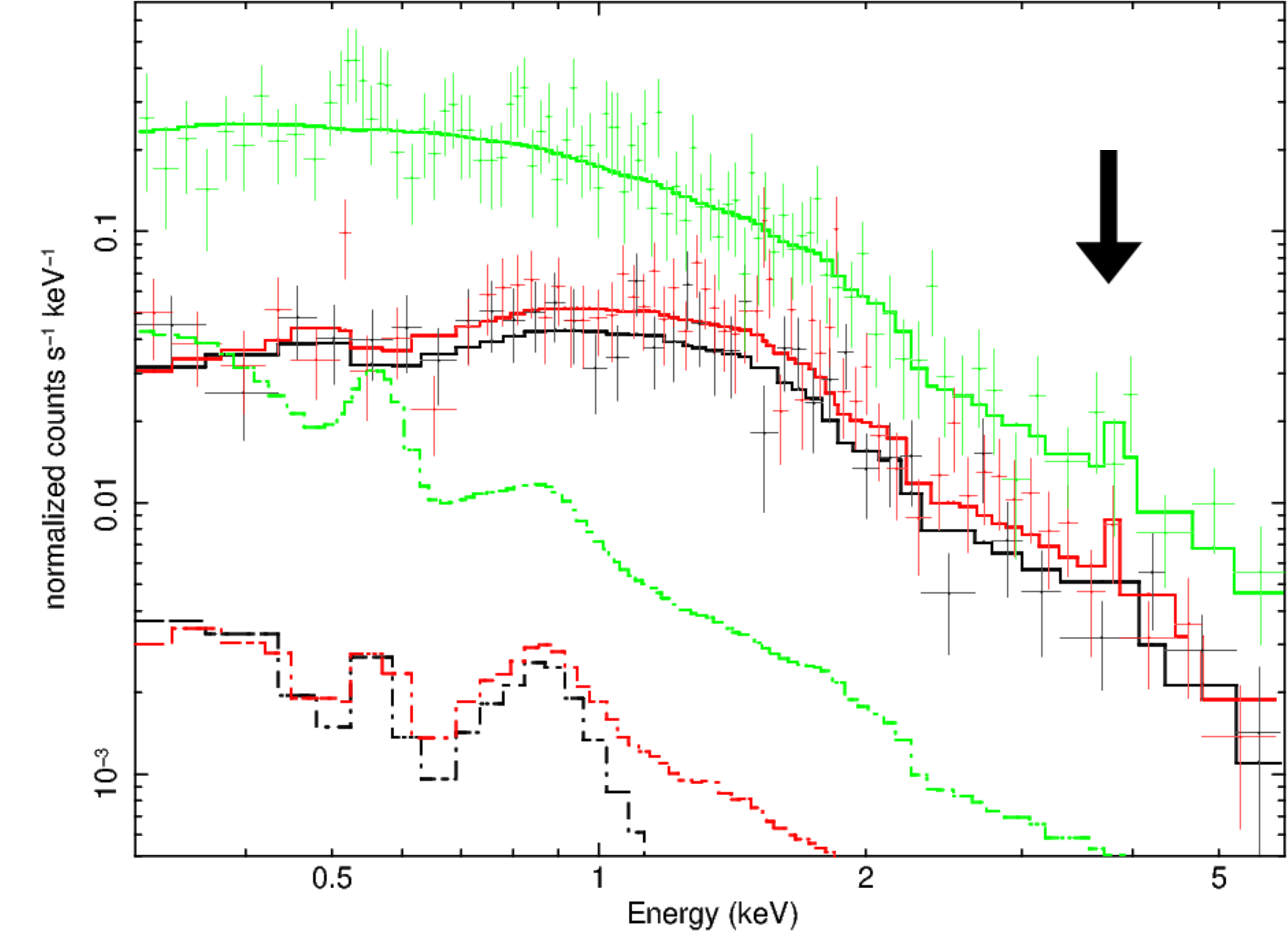}
\end{center}
\caption{\footnotesize{Particle background subtracted spectra extracted from the circular region centred on the X-ray peak of the cluster which radius is defined to maximise the signal-to-noise ratio. The black, red, and green points represent the spectrum extracted from the MOS1, MOS2, and \textit{pn} camera, respectively. The solid lines represent the model used to fit the spectra and their colour legenda is the same as the points. The dash-dotted lines represent the sky background component. The arrow highlights the position of the iron line used to constrain the redshift.}}
\label{fig:spectrum}
\end{figure}
The sky background is estimated modelling the emission of the sky in a region free from cluster emission with two unabsorbed APEC thermal models plus an
absorbed power law with fixed slope of $\Gamma$ = 1.42. The best-fitting model is re-normalised by the ratio of the extraction
areas and added as an extra component in the fit. We then fit simultaneously the instrumental background subtracted spectrum of each camera using an absorbed APEC model plus the sky re-normalised model. The parameters that are free in the fit procedure are the normalisation, the temperature and the redshift of the APEC model. The absorption is folded in using the absorption cross-section of \cite{morrison1983} and fixing the Hydrogen column density to $N_H =2.13 \times 10^{20} cm^{-2}$ as determined from the \cite{kalberla2005} survey. The abundance is fixed to 0.3. 
The fit of the redshift yields $z=0.76^{+0.03}_{-0.04}$, which is in good agreement with the photometric redshift $z_{phot}=0.75 \pm 0.047$. We use this result for all the analysis reported.
The result of the fit is shown in \figiac{fig:spectrum}. Each camera spectrum is shown with different colours and the corresponding model comprising the cluster emission and the sky background is shown with a solid model. The line is visible and its position is highlighted by the black arrow. 

With the X-ray determination of the redshift and the determination of the X-ray peak in hand we can investigate the detection of the joint X-ray SZ COMPRASS catalogue at z=0.5. This detection is probably due to the presence of another cluster at z=0.52 detected by RedMapper \citep{redmapper15} which is $\sim 4$ arcmin distant from \ammasso, the uncertainty of \planck\ position being assumed to be $5$ arcmin.

\subsection{Radial analysis}
The radial density profile of the ICM is measured following the scheme detailed in Sections 3.2 and 3.3 of \cite{bartalucci17}. Firstly, we extracted the instrumental background subtracted and vignetted corrected surface brightness profiles, $S_X$, from concentric
annuli of width $2''$, centred on the X-ray peak. The mean value of the sky background is estimated in a region free of cluster emission and then subtracted. The profile was re-binned to have at least $3\sigma$ in each bin. The $S_X$ profile was used to derive
the radial density profiles, $n_e(r)$, by employing the deprojection technique detailed in \cite{croston2006}. We corrected for the PSF using the model of \cite{ghizzardi2001} who adopts a King function to model the PSF profile as a function of energy and offsets which parameters are reported in EPIC-MCT-TN-011\footnote{\url{http://www.iasf-milano.inaf.it/~simona/pub/EPIC-MCT/EPIC-MCT-TN-011.pdf}} and EPIC-MCT-TN-012\footnote{\url{http://www.iasf-milano.inaf.it/~simona/pub/EPIC-MCT/EPIC-MCT-TN-012.pdf}} for the MOS and pn cameras, respectively; this model has been demonstrated to account for the \xmm\ PSF up to 7 arcsec by \cite{bartalucci17}.

\begin{figure}[!t]
\begin{center}
\includegraphics[width=1\columnwidth]{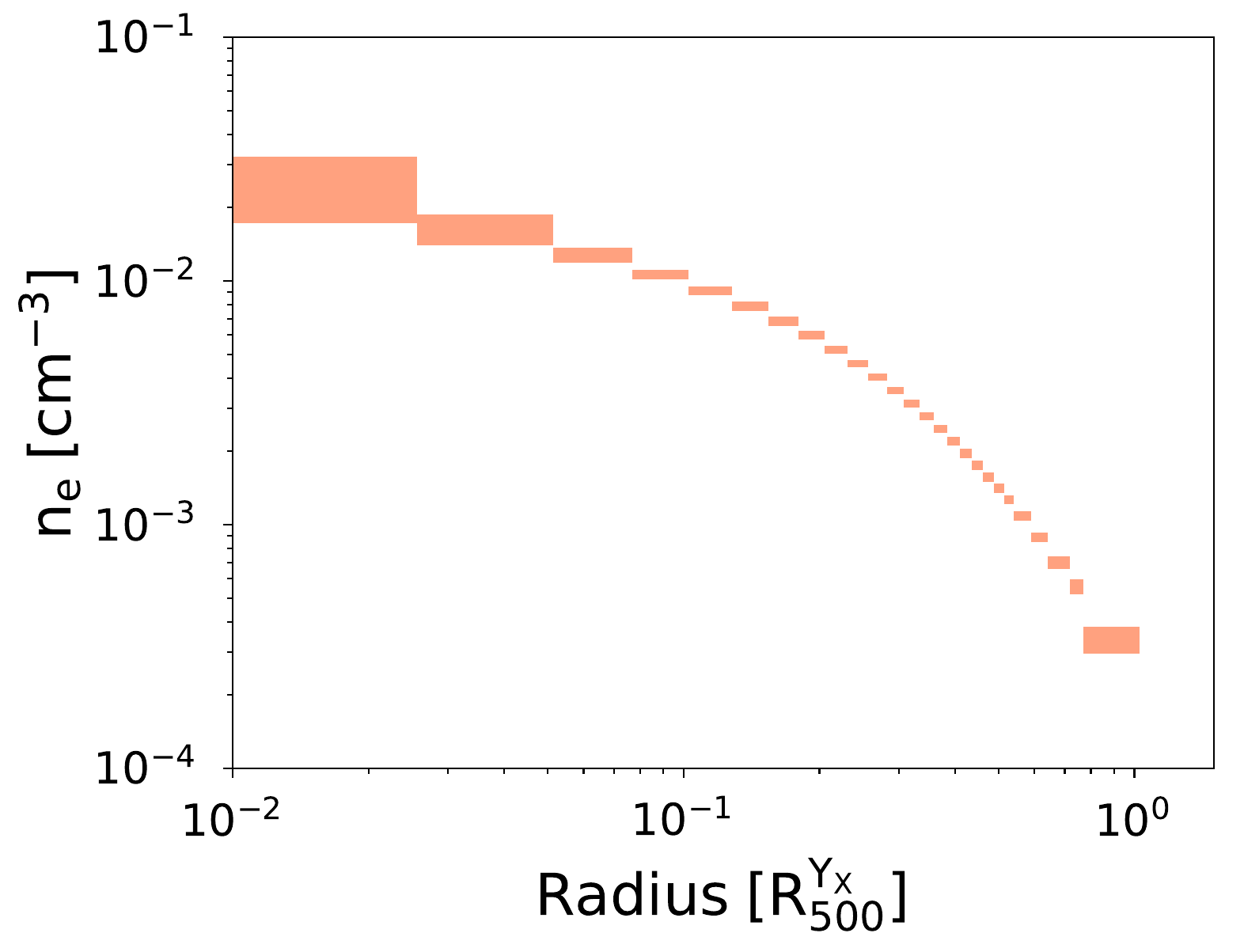}
\end{center}
\caption{\footnotesize{Density profile extracted using as center the X-ray peak and scaled by $R_{500}^{Y_X}$. The size of each polygon along the y-axis represents the $1\sigma$ error.}}
\label{fig:density_prof}
\end{figure}

The scaled density profile of \ammasso\ is shown in \figiac{fig:density_prof} with red rectangles. 
The profile shows no hints of features related to merging activities. This is coherent with the picture emerging from the morphological analysis in which the cluster seems to be mostly dynamically relaxed. The hint of merging phenomena shown above that radius in \figiac{fig:clus_image} are too faint to be detected in the density profile. 

\subsection{Global quantities}\label{sec:global}
The observation is too shallow to extract the temperature radial profile. For this reason, we are not able to measure the hydrostatic mass profile. The measurement of the mass at the density contrast $\Delta = 500$
 with high-precision is fundamental to build a sample of high redshift and massive objects. To do this, we determined the mass, $\Mvyx$, and the corresponding radius, $\Rvyx$, from the mass proxy $\YX$. This quantity is defined as the product of the temperature measured within $[0.15-0.75]\Rv$ region and the gas mass within $\Rv$, as detailed in \cite{krav2006}. We used the \MY\ relation as calibrated from \cite{arnaud2010} assuming self-similar evolution and the gas mass being computed from the volume integration of the density profile. 
 
\begin{table}
\caption{ {\footnotesize Observational and global properties of the cluster analysed in this work.}}\label{tab:clus_prop}
\begin{center}
\begin{tabular}{lc}
\hline        
\hline
RA-DEC (X-peak)  &  $162.0123$; $31.6449$ [J2000]\\[0.08cm]
Redshift$^a$   & $0.76^{+0.03}_{-0.04}$ \\[0.08cm]
$\Mvyx{}$   & $5.64^{+0.79}_{-0.62} \times 10^{14}$ $[M_{\odot}]$\\[0.08cm]
$\Rvyx$ $^b$    & $950^{+43}_{-36}$ [kpc]\\[0.08cm]
$T_{\textrm X}$ $^c$ & $7.0^{+1.1}_{-0.9}$ [keV]\\[0.08cm]
$M_{gas}(<\Rvyx)$ & $8.2^{+0.8}_{-0.6} \times 10^{13}$ $[M_{\odot}]$\\[0.08cm]
\hline
\end{tabular}
\end{center}
\footnotesize{Notes: $^{(a)}$ Redshift determined from X-ray spectroscopy as described in \ref{sec:redshift}. $^{(b)}$ The radius in arcmin is $2.16^{+0.10}_{-0.08}$. $^{(c)}$ Temperature measured from the fit of the X-ray spectrum extracted from the [0.15-0.75]$\Rvyx$ circular region.}
\end{table}
The results of this computation as well as other global quantities are summarised in Table \ref{tab:clus_prop}. The $\Mvyx = 5.64^{+0.79}_{-0.62} \times 10^{14} [M_{\odot}]$ is consistent at $1\sigma$ with the value of $M_{500}=9.8\pm 3.2 \times 10^{14} M_{\odot}$ computed by \cite{clus_candiate15} through the \MYSZ\ relation but they differ by almost a factor 2. This is not surprising, the scatter of this relation being already  of the order of $\sim 20-30\%$. 

\section{Discussion}

We presented in this work the X-ray analysis of \ammasso\ leveraging its serendipitous observation. With this \xmm\ observation in hand we were able to:
\begin{itemize}
    \item investigate the morphology of the cluster within $\Rvyx$ and infer the dynamical status which appears to be relaxed in the inner part with hints of interacting substructures in the outskirts. However, the lack of significant merger features could simply be due to the result low statistics combined with low angular resolution of the data.  The cluster does not appear to host a cool core;
    \item confirm the optical photometric redshift and benchmark the possibility of using X-rays to estimate the redshift at such low-statistic regime. This results shows the efficiency of \xmm\ snapshots in being useful not only to confirm cluster presence but also give important information such as redshift;
    \item measure the density profile up to $\Rvyx$. This quantity strengthens the picture emerging from the morphological analysis;
    \item combine the gas density profile and temperature measured within fixed apertures to estimate the mass through the low-scatter and high-precision mass proxy $Y_X$, which yields an unprecedented $\sim 10\%$ precision measurement of the mass. 
\end{itemize}
We stress the fact that we were able to achieve such level of characterisation with a short-exposure observation and, furthermore, the object of interest is 9 arcmin offset from the aimpoint. 

\begin{figure}[!t]
\begin{center}
\includegraphics[width=1\columnwidth]{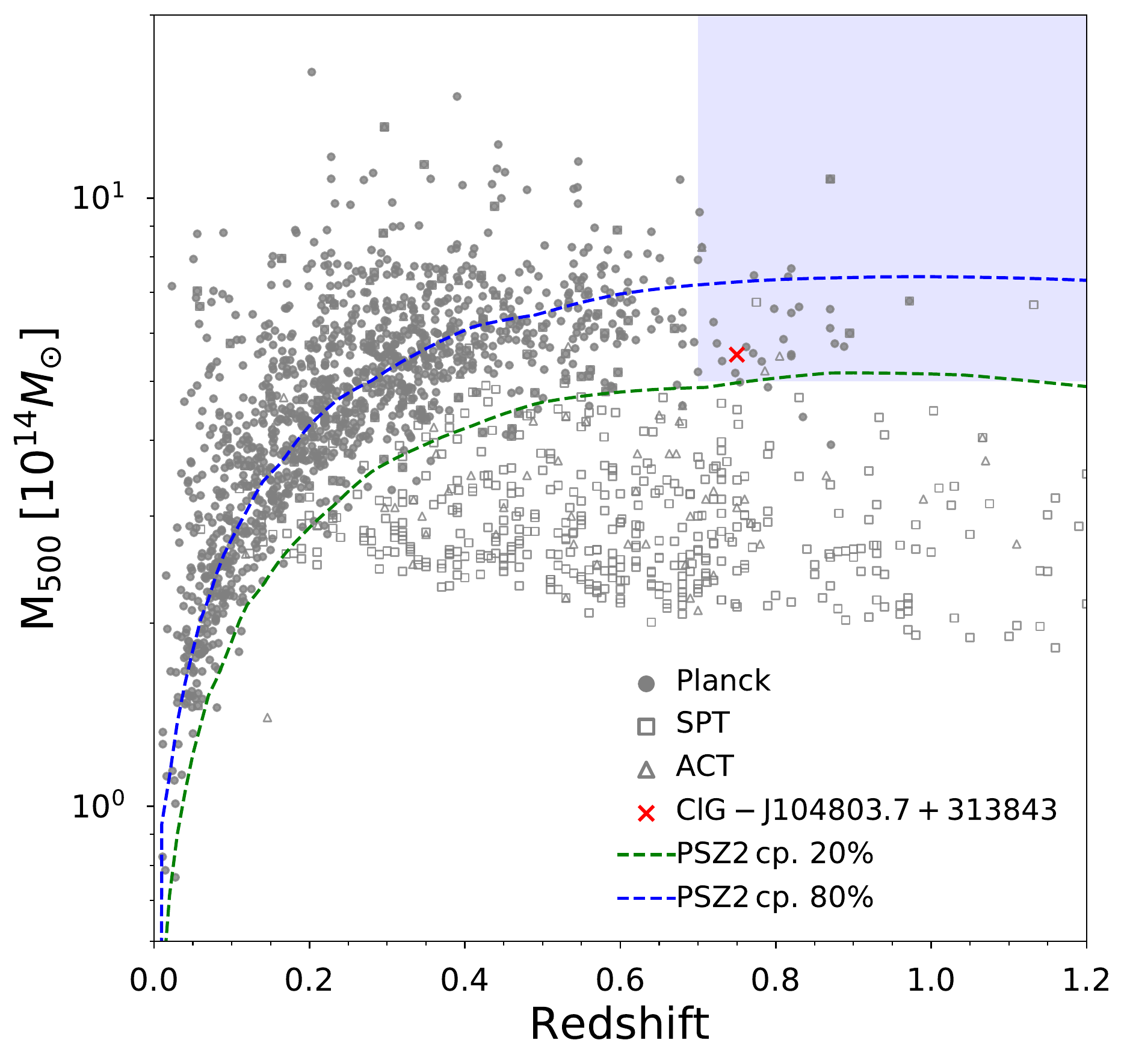}
\end{center}
\caption{\footnotesize{Cluster distribution in the mass-redshift plane. Filled points: clusters identified by the \textit{Planck} all-sky survey; empty boxes: clusters from the SPT collaboration; empty triangles: clusters from the ACT survey. Masses in the \textit{Planck} catalogue are derived iteratively from the \MYSZ\ relation calibrated using hydrostatic masses
from \xmm. They are not corrected for the hydrostatic
equilibrium (HE) bias. Published SPT masses are estimated ‘true’
mass and are re-normalised by a factor 0.8 to the \planck\ standard
for comparison. The confirmed cluster studied in this work is shown using a red cross. The blue shaded
box identifies a region of massive clusters, $M_{500} > 5 \times 10^{14} M_{\odot}$, at $z>0.7$ which is poorly populated. There are $\sim 25$ objects in this region as found by all the SZ surveys. The dashed green and blue lines indicate the 20\% and 80\% survey completeness contours, respectively, of the \planck\ SZ release 2 (PSZ2) catalogue which are shown in Fig. 26 of \cite{PSZ2}. }}
\label{fig:planck_plot}
\end{figure}

The building of a well-characterised sample of high-redshift and massive clusters is a crucial point for any study envisaging to significantly improve our understanding of these peculiar objects. We show in \figiac{fig:planck_plot} the distribution in the mass-redshift plane of all the clusters found by the \textit{Planck}, SPT, and ACT SZ surveys. The \ammasso\ cluster is shown with a red cross. The region of the most massive and distant redshift objects highlighted with a blue polygon is the less populated region as compared to others, and any effort to add even one cluster is fundamental. Optical based survey are also fundamental to enrich these populations.  MadCows \citep{gonzalez19} successfully delivered  $\sim 2000$ candidates at $z>0.7$. However, the mass $\Mv$ measured with a low-scatter proxy such as the $Y_{SZ}$ is available only for 14 clusters, only 3 being more massive than $\Mv > 5\times10^{14} M_{\odot}$.

The multiwavelength approach is the game-changer to study high redshift and massive objects. The dashed coloured lines shown in \figiac{fig:planck_plot} represent the minimum mass above which we have a probability of $20\%$ and $80\%$, in green and blue, respectively, to find  a cluster with \planck\ i.e. the completeness of the sample at a given redshift and mass. These curves  show that \planck\ alone is not capable of delivering numerous massive and distant objects. The current combination of relatively limited sensitivity at high redshift of the all-sky \planck\ survey and the limited sky coverage of the ground-based SZ observatory will leave to the future combination of optical-Near-infrared facilities such as LSST and Euclid and to eROSITA to consistently probe the population of the massive clusters at high redshift e.g. \cite{mantz19} and references therein. X-rays and their combination with the new SZ high resolution observations are the key elements to derive fundamental thermodynamic quantities such as density and temperature or dynamical information such as the hydrostatic mass profile.

Furthermore, the characterisation of this sample is fundamental also to carefully plan future observation campaigns with deep observations. That is, observing these objects is extremely time-consuming and knowing in advance the mass and the morphology is crucial to carefully select objects. \ammasso\ has been initially detected as a $\sim 10^{15} M_{\odot}$ object while our work shows that is still massive but has a smaller mass with a reduction on the relative error of the order of $\sim 3$ times. This result shows the importance of the X-ray characterisation.

\begin{acknowledgements} 
The results reported in this article are based on data obtained from the \xmm\ observatory, an ESA science mission with instruments and contributions directly funded by ESA Member States and NASA.  EP and LZ acknowledge financial support under ASI/INAF contract 2017-14-H.0. ML acknowledges financial support from the Ph.D. programme in Astronomy, Astrophysics and Space Science supported by MIUR (Ministero dell’Istruzione, dell’Università e della Ricerca).

 \end{acknowledgements}

\bibliographystyle{aa}
\bibliography{lib_articoli}
\appendix

\end{document}